# A universal law for the pattern evolution of fullerene-based sandwiches

Yixuan Xue[a], Jin-Wu Jiang[a,*], Harold S. Park[b]

[a]Shanghai Key Laboratory of Mechanics in Energy Engineering, Shanghai Institute of Applied Mathematics and Mechanics, School of Mechanics and Engineering Science, Shanghai University, Shanghai 200072, People's Republic of China
[b]Department of Mechanical Engineering, Boston University, Boston, Massachusetts 02215, USA

**Abstract**

Fullerene-based sandwiches have emerged as new candidates for potential applications of two-dimensional nanomaterials in electronics or energy storage. Recently, experimentalists have observed the evolution of boundaries for fullerene clusters sandwiched by two graphene layers, while vacuum space with typical dimension of 30 Å was found within the fullerene layer. Because the pattern of the fullerene cluster impacts the physical properties of the sandwiches, it is important to understand the mechanisms for their structural transformations. In the present work, we find that the graphene/fullerene/graphene sandwich structure transforms among three configurations, depending on the fullerene to graphene area ratio. Molecular dynamics simulations show that there are two critical values for the area ratio. The fullerene pattern transforms from circular to rectangular at the first critical area ratio of $\frac{1}{\pi}$. The critical value of $\frac{1}{\pi}$ is successfully derived by comparing the geometrical perimeter of the circular and rectangular shapes for the fullerene cluster without any physical parameters. At the second critical area ratio, the graphene layers surrounding the fullerene cluster become separated due to the competition between the bending energy and the cohesive energy. Based on the analytic model, the vacuum space is predicted to be 34 Å, which agrees quite well with the experimental result. These findings provide fundamental insights into the mechanisms driving structural transformation of fullerene-based sandwiches, which will enable the selection of appropriate structures and materials to guide the design of future electronic and energy storage applications.

## 1. Introduction

Sandwich-like structures can be formed by the stacking of atomic layers in a layer-by-layer mode. Among others, fullerene-based sandwiches significantly extend the functionalities and applications of two-dimensional (2D) nanomaterials by leveraging the distinct properties of fullerene and 2D nanomaterials[1–8]. For example, the fullerene and graphene oxide composites have high solubility and good stability. Fullerenes can be supported on graphene oxide and dissolved in water with the aid of the hydrophilic groups of graphene oxide, because of the π-stacking interaction between the surface of graphene oxide and the sidewall of fullerene[3]. The fullerene/MoS$_2$ hybrid structure shows abundant electronic properties, such as sustainable conductivity, bistability and typical bipolar resistance switching phenomenon with low set and reset voltages, and high ON/OFF resistance ratio ($4 \times 10^3$), which are not possible in the individual fullerene or MoS$_2$ constituents[5].

The graphene/fullerenes/graphene (GFG) sandwiches have been synthesized by Mirzayev et al. in a recent experiment[9]. The experiment observed a clear boundary for the fullerene cluster sandwiched by graphene layers, and the oscillation of fullerenes between different positions. In the experiment, several stable configurations are observed for the GFG sandwich structure. There are two major findings for the structure of the GFG sandwich structure. First, the boundary of the fullerene cluster can be self-assembled into different stable shapes. Secondly, experimental measurements found microscopic vacuum space of about 30 Å can be formed within the fullerene layer that is sandwiched by graphene layers. Furthermore, the fullerene clusters sandwiched by other 2D nanomaterials also have different stable patterns as the number of fullerenes changed[10–17].

The pattern of the cluster can affect properties for the GFG sandwich structure. For example, a promising approach to modulate the bandstructure of the graphene is the modification of graphene's topography using fullerenes[18, 19]. The cognate carbon nanostructures sandwiched by graphene layers can modify the topography of the wrinkle network[20]. In addition to the topography-induced changes, the fullerene structure arrangement may result in a significant modification of some physical properties. Buckyball (C$_{60}$) clusters were sandwiched between graphene layers to tune the cross-plane thermal conductivity in few-layer graphene[21]. When the C$_{60}$ clusters are wrapped by graphene, the cross-plane thermal conductivity increases with the increase of vdW interactions. Ojeda-Aristizabal et. al found that charge transfer between fullerene and the graphene is sensitive to the nature of the underlying supporting substrate and to the crystallinity and local orientation of the fullerene[22]. Although the pattern of the fullerene cluster is important, there is still no analytic model for the evolution of the pattern for the fullerene cluster sandwiched by atomic layers. Therefore, it is an important task to study the patterns in the sandwich structure and understand the microscopic mechanism for the pattern evolution, which would enable the tailoring of the sandwich physical properties.

In this paper, the pattern for the fullerene layer sandwiched by two graphene layers is studied by molecular dynamics (MD) simulations and analytic models. The configuration of the GFG sandwich structure is determined by the fullerene to graphene area ratio. More specifically, there are two critical values for the area ratio, which govern the transition of the GFG sandwich structure among three configurations. For large systems, the pattern of the fullerene layer transforms from circular to rectangular at the critical area ratio of $\frac{1}{\pi}$, which can be analytically derived by comparing the geometrical perimeter of these two patterns. At the second area ratio, the graphene layers surrounding the fullerene layer transforms from the adhered configuration into the separated configuration, which can be analyzed by theoretical models based on the competition between the bending energy and the cohesive energy. Predictions from the analytic models are compared with available experiments and good agreement is achieved. The analytic models are universal and are applicable to other sandwich structures such as the $MoS_2$/fullerene/$MoS_2$ sandwich structure beside the GFG structure.

## 2. Structure and simulation details

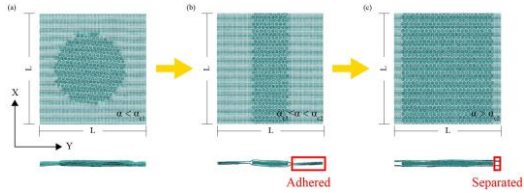

Figure 1: Three typical configurations of the GFG sandwich structure with the increase of the ratio α. Top and bottom panels are the top and side views for the sandwich structure.

Fullerenes can be sandwiched by two atomic layers. Fig. 1 shows that the $C_{60}$ fullerene layer is sandwiched by two graphene layers, forming the GFG sandwich structure. To demonstrate the universality of the results, the $MoS_2$/fullerenes/$MoS_2$ sandwich structure will also be studied. Fig. 1 shows the evolution of the pattern for the middle self-assembled fullerene layer in the GFG sandwich structure with the increase of the structural ratio α. Here α is the area ratio, $\alpha = \frac{A}{A_0}$, where $A$ is the area of the fullerene pattern and $A_0 = L^2$ is the area of the square graphene layer. From Fig. 1 (a) to Fig. 1 (b), the pattern changes from a circle into a rectangle, in which the two graphene layers adhere together. In Fig. 1 (c), the fullerenes are in the rectangular pattern, while the graphene layers surrounding the fullerene cluster are separated. In this work, we will show that there are two critical value $\alpha_{c1}$ and $\alpha_{c2}$ for the evolution of the pattern for the fullerene sandwiched by two atomic layers. The pattern evolves from Fig. 1 (a) to Fig. 1 (b) at α = $\alpha_{c1}$, while the pattern changes from Fig. 1 (b) to Fig. 1 (c) at α = $\alpha_{c2}$.

MD simulations are carried out by using the Large-scale Atomic/Molecular Massively Parallel Simulator (LAMMPS) package[23]. Newton's equations of motion are integrated by the velocity Verlet algorithm with a time step of 1 fs[24]. Periodic boundary conditions are applied in the in-plane X and Y directions, while free boundary conditions are applied in the out-of-plane Z direction. The adaptive intermolecular reactive empirical bond-order (AIREBO) potential is used to describe the covalent and vdW interactions of the molecular crystal structure for the GFG sandwich structure[25]. The Lennard-Jones (LJ) cutoff between all carbon atoms for AIREBO potential is considered to be 10 Å. Simulations are performed in the isothermic-isobaric (NPT) ensemble at room temperature and zero pressure. The graphene sheets are relaxed for 100 ps, and then the sandwich structure is thermalized for 400 ps. All figures for the atomic structures are created by VMD[26]. We note that there is some mismatch strain between the graphene sheets and the fullerene layer. The dimensions of the system are properly chosen so that the mismatch strain is less than 1 %.

## 3. Circular versus rectangular pattern

### 3.1. MD results

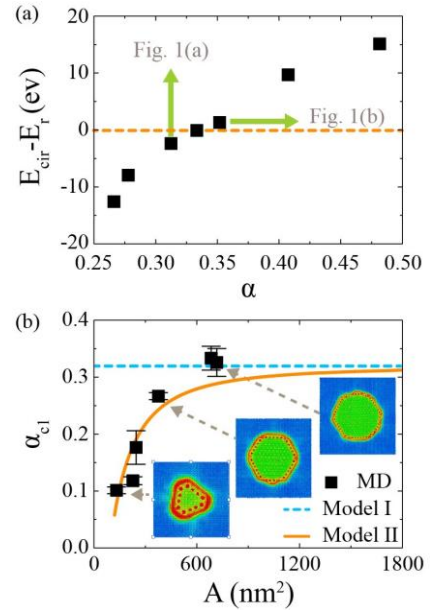

Figure 2: The structure dependence for the critical value $\alpha_{c1}$. (a) The energy difference, $E_{cir} - E_r$, increases with increasing ratio α and changes its sign at a critical value $\alpha_{c1}$ for a given graphene area, where $E_{cir}$ is the potential energy of the sandwiches with circular fullerene pattern and $E_r$ is the potential energy with rectangular fullerene pattern. (b) The critical value $α_{c1}$ increases with the increase of graphene area, and saturates at a constant value of $\frac{1}{π}$. Insets show different shapes of the fullerene layer, where the color bar represents the atomic potential energy.

We perform MD simulations to compare the potential energy for the GFG sandwich structure with circular and rectangular fullerene clusters. For a given graphene bilayer, we increase the ratio α by increasing the number of fullerenes. For a given number of fullerenes (with a specific value of α), the fullerene layer can be either in the circular pattern or the rectangular pattern. To compare the stability of these two patterns, we predesign the GFG sandwich structure in the circular pattern or the rectangular pattern. We then perform MD simulations for both patterns. Note that these patterns will not transform from one into the other in the MD simulation, as the energy barrier between these two patterns is so high that the fullerenes can be stationary in a particular shape for as long as 1 ms[9]. For the GFG sandwich structure with different ratio α, we record the energy difference, $E_{cir} - E_r$, where $E_{cir}$ is the potential energy of the sandwiches with circular fullerene layer and $E_r$ is that for the rectangular fullerene pattern. Fig. 2 (a) shows that with the increase of the ratio α, the sign of the energy difference changes from negative to positive at a critical value $α_{c1}$. It indicates that for a given graphene bilayer, the most stable pattern of the fullerene layer will evolve from circular to rectangular at a critical ratio $α_{c1}$ with increasing fullerene number.

Figure 2 (b) shows that the critical value $α_{c1}$ is dependent on the size of the graphene layer. The critical value $α_{c1}$ increases with increasing size of the graphene layer, and saturates at a constant value of $\frac{1}{π}$ in the limit of large graphene layer. For each structure, we perform five MD simulations with different initial random velocity distribution. These five data are averaged to give the error bar in the MD simulation result.

*3.2. Analytic models*

We now develop analytic models to understand the underlying mechanism for the above MD results on the transition from the circular pattern to the rectangular pattern at the critical ratio $α_{c1}$. From the energy point of view, the configuration of the GFG sandwich structure is determined by two typical energies, i.e., the cohesive energy and the bending energy. The cohesive energy is the coupling between adjacent atomic layers, including the cohesion between the fullerene layer and the graphene layer within the fullerene region, and the cohesion between graphene layers surrounding the fullerene region. The bending energy is caused by the deformation of the graphene layer.

*3.2.1. Model I*

The number of fullerenes is the same for the circular and rectangular patterns in Figs. 1 (a) and (b), so the area of the fullerene region is the same in these two patterns. The cohesive energy between the fullerene layer and the graphene layer is proportional to the area of the fullerene region, so the cohesive energy between the fullerene layer and the graphene layer is the same in the circular and rectangular patterns. Similarly, the cohesive energy between adjacent graphene layers is the same in the circular and rectangular patterns. As a result, the total cohesive energy is the same in the circular and rectangular patterns in Fig. 1 (a) and (b).

The bending energy describes the deformation of the graphene layer. The bending deformation of graphene occurs only at the boundary of the fullerene pattern (see the red region in Fig. 3 (a)), so the bending energy is proportional to the length of the edge for the fullerene pattern. As a result, the bending energy is

$$E_b = ρ_b × L_{edge}, \qquad (1)$$

where $ρ_b$ is the bending energy density per length and $L_{edge}$ is the length of the edge for the fullerene pattern. The bending energy density $ρ_b$ is the same in the circular and rectangular patterns. It should be noted that the present model is for isotropic materials. For anisotropic materials like black phosphorous, the bending energy density depends on the bending direction, so the bending energy will depend on the pattern type. The critical value $α_{c1}$ occurs when the bending energy for these two patterns equals to each other, i.e., the perimeter of these two patterns equals

$$2πr = 2\sqrt{A_0}. \qquad (2)$$

Note that there are only two edges in the rectangular pattern due to the periodic boundary condition along the in-plane y direction. The area for the circular pattern is $A = πr^2$, so we have $r = \sqrt{\frac{A}{π}}$. Inserting this expression into Eq. (2), we get the critical value for the ratio,

$$α_{c1} = \frac{A}{A_0} = \frac{1}{π}. \qquad (3)$$

The analytic prediction of $α_{c1} = \frac{1}{π}$ is plotted as the horizontal line in Fig. 2 (b). We find that the MD results saturates to this analytic prediction in the limit of large graphene sheets. It should be noted that the analytic result in Eq. (3) is derived purely based on the geometry of the fullerene pattern, without any physical parameters. This formula is thus universal and should be applicable to any atomic layer and any type of fullerenes.

Figure 2 (b) also shows that the critical value $α_{c1}$ depends on the size of graphene layer for small pieces of graphene. The analytic result in Eq. (3) cannot capture this size effect on the critical ratio $α_{c1}$, due to the extreme simplicity of the analytic model. We will show next that this size effect can be included by considering some details for the shape of the pattern.

*3.2.2. Model II*

In the previous simple model, we have assumed that the pattern is a perfect circle for small values of the ratio α. The pattern transfers into the rectangular shape for $α > α_{c1}$. The bending energy can be minimized in the circular shape, which has the smallest perimeter for a given number of fullerenes. It is thus reasonable to assume a circular pattern. However, the shape of the pattern can deviate from a circle when the number of fullerenes is very small (small α). The insets of Fig. 2 (b) display the actual evolution of the shape for the fullerene pattern with increasing fullerene number. For small number of fullerenes, the pattern is a polygon, and the number of the polygonal edges increases gradually with increasing dimension for graphene. We thus use a polygonal shape to substitute the circular shape in Eq. (2). For a polygon of area $A$, the perimeter is

$$L_{edge} = M \times 2R \times \sin\frac{\pi}{M}, \quad (4)$$

where $M$ is the number of the edge for the polygon, and $R = \sqrt{2A/(M\sin\frac{2\pi}{M})}$ is the distance from the center to the apex for the polygon. By equating the perimeter of the polygon and the rectangle, we get

$$\alpha_{c1} = \frac{1}{M \times \tan\frac{\pi}{M}}. \quad (5)$$

The insets in Fig. 2 (b) show that the number of the polygonal edges increase with increasing graphene area $A_0$. We make the simplest linear assumption for the polygon, $M = aA_0 + b$, where $a$ and $b$ are fitting parameters. The model in Eq. (5) is plotted as the orange line in Fig. 2 (b) with fitting parameters $a$ = 0.0064 and $b$ = 1.4054. We find that the improved model qualitatively agrees with the MD results in the full range of the graphene area.

*3.3. Possible experiments*

Presently, experiments have not been performed to examine the transition from the circular pattern into the rectangular pattern. However, current experimental techniques are able to manipulate the size of the fullerene cluster and the graphene dimension in the GFG sandwich structure[9]. The detailed geometry for the fullerene pattern can be accurately detected by the scanning electronic microscopic (SEM) set up, which has been applied to monitor the evolution of the edge for the fullerene pattern[9]. Hence, we believe that the MD results and the universal analytic formula discussed in this section are readily to be verified experimentally.

**4. Adhered structure versus separated structure**

In the previous section, we have shown that the fullerene pattern transforms from circular to rectangular at $\alpha_{c1}$ with increasing number of fullerenes. The fullerenes are sandwiched by the graphene layers, where the top and bottom graphene layers are adhered with each other in these regions surrounding the fullerene. With further increasing fullerene number, it is possible that these two adhered graphene layers will be separated as illustrated by Fig. 1 (c). We will now study this structural transition.

*4.1. MD results*

Figure 4 shows the relation between the critical ratio $\alpha_{c2}$ and the length of the structure. The structure is a square and the size of graphene increases from 9.6 nm to 33.6 nm for this set of simulations. The critical ratio $\alpha_{c2}$ is larger than $\alpha_{c1}$. The fullerene is in the rectangular pattern for these structures simulated for Fig. 4, which is the most stable pattern for $\alpha > \alpha_{c1}$. Here $\alpha_{c2}$ is the critical ratio, at which the adhered graphene layers are separated. We find that the critical value $\alpha_{c2}$ is dependent on the structure size. More specifically, $\alpha_{c2}$ increases with increasing size and approaches to 1.0 in the limit of large system size.

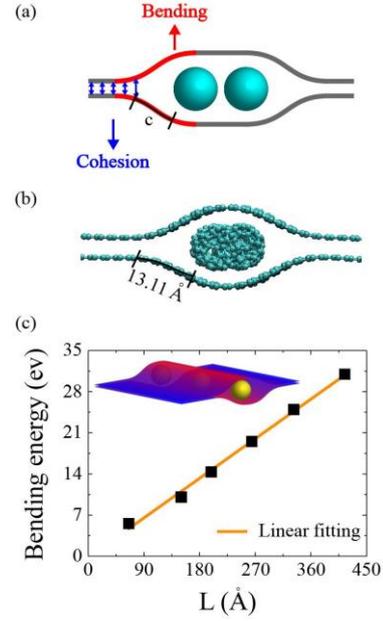

Figure 3: Local configuration for the GFG sandwich structure. (a) Sketch structure. (b) Snapshot from MD simulation. (c) Bending energy density per length. The bending energy versus the length $L$. The data is linearly fitted to $E_b$ = 0.075$L$, giving the bending energy density per length as $\rho_b$ = 0.075 eV/Å. The illustration is the schematic diagram for the energy distribution.

*4.2. Analytic models*

We develop analytic models to describe the structural transition from Fig. 1 (b) to Fig. 1 (c) at $\alpha_{c2}$. For a given sandwich structure, we compare the potential energy for these two configurations shown in Fig. 1 (b) and (c). The structure prefers to stay in the configuration with lower potential energy. In the first configuration as shown in Fig. 1 (b), the top and bottom graphene layers surrounding the fullerenes are adhered, which we call the adhered configuration. In the second configuration as shown in Fig. 1 (c), the top and bottom graphene layers surrounding the fullerenes are separated, which will be referred to as the separated configuration. We will propose two models, which reflect the competition between bending and cohesive energies, to compute the potential energy for these two configurations.

The bending energy density per length is calculated in Fig. 3 (c). We calculate the bending energy in the graphene bilayer sandwiching an array of fullerenes, where the length of the fullerene array increases from 50 Å to 400 Å. A linear fitting in Fig. 3 (c) gives the bending energy density per length to be $\rho_b$ = 0.075 eV/Å. The cohesive energy density per area $g_c$ = −0.019 eV/Å$^2$ is calculated from the cohesive energy for a bilayer graphene.

*4.2.1. Model III*

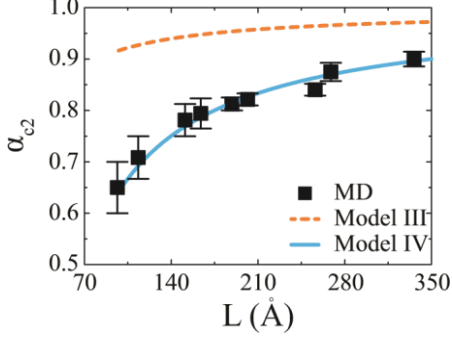

Figure 4: The critical value $\alpha_{c2}$ increases with the increase of the structure dimension $L$.

For the adhered configuration, the bending energy is proportional to the perimeter of the fullerene pattern,

$$E_b = 2L\rho_b, \qquad (6)$$

where $L$ is the size of the square sandwich structure. The cohesive energy within the bilayer graphene surrounding the fullerene pattern is

$$E_c = g_c(A_0 - A) = g_c L^2 (1-\alpha), \qquad (7)$$

where $A_0 = L^2$ is the area for the whole square graphene, and $A$ is the area for the fullerene region.

For the separated configuration, the bending energy is zero, because both graphene layers are in the planar configuration without bending. The cohesive energy within the bilayer graphene surrounding the fullerene pattern is also zero, as these two graphene layers are separated from each other. Note that there is also cohesive energy between the graphene layer and the fullerene layer, which is not considered since this cohesive energy term is the same in the adhered and separated configurations.

As a result, the transition from the adhered configuration to the separated configuration occurs when the total potential energy for these two configurations balance each other, i.e., $E_b + E_c = 0$, so we have

$$2L\rho_b + g_c L^2(1-\alpha_{c2}) = 0. \qquad (8)$$

From Eq. (8) we get

$$\alpha_{c2} = 1 - \frac{2\rho_b}{|g_c|L} \qquad (9)$$

The analytic result in Eq. (9) for model III is shown in Fig. 4. The predicted $\alpha_{c2}$ increases with increasing system size, which qualitatively agrees with the MD simulation results. However, it can be seen that the model overestimates the critical value $\alpha_{c2}$.

*4.2.2. Model IV*

In Model III, the cohesive energy is computed in a rather simple manner. We need to carefully examine the boundary regions surrounding the fullerene pattern (see the red region in Fig. 3 (a)). In Model III, the cohesive energy is calculated for all regions except the fullerene region. In particular, the bending graphene segment surrounding the fullerene region is also assumed to be perfectly adhered. However, the top and bottom bending graphene segments of length $c$ actually do not adhere with each other. This is the origin for the overestimation for $\alpha_{c2}$ of Model III. Hence, we propose to correct the cohesive energy for the adhered configuration,

$$E_c = g_c(A_0 - A - 2Lc), \qquad (10)$$

where $c$ is a parameter describing the characteristic length for the bending segment surrounding the fullerene pattern as shown in Fig. 3 (a). The bending energy is the same as Eq.(6) in Model III. From Eq.(6) and Eq.(10), we get the critical ratio,

$$\alpha_{c2} = 1 - \frac{2\rho_b}{|g_c|L} - \frac{2c}{L}. \qquad (11)$$

This analytic formula is plotted in Fig. 4, where the fitting parameter is $c$ = 13.11 Å. We obtain an excellent agreement between the analytic result and the MD simulation results. The characteristic length $c$=13.11 Å is a rather reasonable value as can be seen from the MD snapshot in Fig. 3 (b).

*4.3. Comparison with experiments*

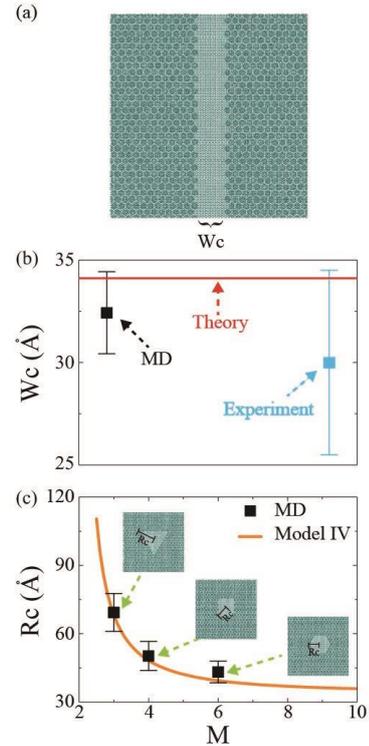

Figure 5: The maximum width $W_c$ for the free space in the GFG sandwich structure. (a) The top view for a rectangular free space. (b) The value for $W_c$ from experiments, MD simulations and the theoretical prediction. (c) The maximum dimension $R_c$ for the polygonal free space in the GFG sandwich structure.

In the structure shown in Fig. 5 (a), fullerenes are sandwiched by the top and bottom graphene layers in the left and right regions. The central graphene layers of width $W_c$ are separated, forming a free space on the nanometer scale. This free space is also found in experiments[9], and can provide a nanoscale reaction chamber and a clean interface in the microscope vacuum. In the experiment, the width $W_c$ for the free space is about 30 Å. Using the above analytic models, we can predict a theoretical value for $W_c$.

The structure shown in Fig. 5 (a) is essentially the same as Fig. 1 (c). We have shown in the above that there is a critical value $\alpha_{c2}$, above which the free space can exist. As a result, $W_c$ can be extracted from $\alpha_{c2}$ as follows,

$$\alpha_{c2} = \frac{(L - W_c)}{L}, \tag{12}$$

which gives

$$W_c = (1 - \alpha_{c2}) \times L. \tag{13}$$

Using Eq. (11) for $\alpha_{c2}$ in Model IV, we get

$$W_c = \frac{2\rho_b}{|g_c|} + 2c. \tag{14}$$

Inserting all parameters into Eq. (14), we get the analytic value of $W_c$ = 34.11 Å. We also perform MD simulations to calculate $W_c$, and the simulation value is 32.43 Å. These results are plotted in Fig. 5 (b). We find reasonable agreement between our analytic result and the experimental or MD simulation results.

*4.4. Predictions from the analytic model*

We can also predict the critical dimension for the free space of any polygonal shape in the GFG sandwich structure. Insets of Fig. 5 (c) display three shapes for the free space. The top and bottom graphene layers are separated from each other in the polygonal free space. We use the distance $R$ from the center to the apex of the polygon to characterize the size of the polygon. MD results are displayed as points in Fig. 5 (c). It shows that the critical size for the polygonal free space depends on the shape of the polygon. We apply the above Model IV to analyze the MD results. The perimeter of regular polygon with size number $M$ is described by Eq. (4), so the bending energy is $E_b = \rho_b L_{edge}$.

According to Model IV, the adhesive area is

$$S = M \times R^2 \times sin\frac{\pi}{M} \times cos\frac{\pi}{M} - cL \tag{15}$$

where the parameter $c$ = 13.11 Å describing the characteristic length for the bending segment surrounding the fullerene pattern is from Model IV. The cohesive energy is $E_c = g_c S$. From $E_b + E_c = 0$, we get the critical size $R_c$,

$$R_c = \frac{2(\rho_b + cg_c)}{g_c cos\frac{\pi}{M}}. \tag{16}$$

This analytic formula is plotted in Fig. 5 (c). The predicted $R_c$ decreases with increasing $M$, which agrees quite well with the MD simulation result.

All analytic models which describe the pattern evolution of the GFG sandwich structure are universal and are applicable to other fullerene-based sandwiches with isotropic bending energy. We demonstrated it with the $MoS_2$/fullerene/$MoS_2$ structure in the Supplementary material.

**5. Conclusion**

We have performed MD simulations to investigate the mechanics underlying pattern formations for fullerene clusters that are sandwiched by graphene bilayers. The simulations show that as the ratio of fullerene to graphene area α changes, the resulting stable sandwich structure evolves among three typical configurations, which are governed by two critical ratios. At the first critical ratio $\alpha = \alpha_{c1}$, the fullerene pattern changes from circular into rectangular. At the second critical ratio $\alpha = \alpha_{c2}$, the graphene layers surrounding the fullerene cluster are separated by the fullerenes. The MD results can be well explained by analytic models based on the interplay between the bending energy and the cohesive energy, which also successfully explain existing experimental observations. To demonstrate the generality of the analytic models, we have illustrated that the analytic models are applicable to other sandwich structures such as $MoS_2$/fullerene/$MoS_2$ besides the GFG sandwich structure. Because the analytic models are based on a simple energy competition mechanism, similar sandwich structures can also be described by such approaches across different length scales. Thus, we expect that this study can provide guidance for theory and experiment to design fullerene-based sandwiches and related applications.

**Acknowledgment** The work was supported by the National Natural Science Foundation of China (Grant Nos. 11822206 and 12072182), Innovation Program of the Shanghai Municipal Education Commission (Grant No. 2017-01-07-00-09E00019) and Key Research Project of Zhejiang Laboratory (No. 2021PE0AC02).

∗Corresponding author
*Email address:* jwjiang5918@hotmail.com (Jin-Wu Jiang)

# Supplementary Material

# A universal law for the pattern evolution of fullerene-based sandwiches

Yixuan Xue

Shanghai Key Laboratory of Mechanics in Energy Engineering,
Shanghai Institute of Applied Mathematics and Mechanics,
School of Mechanics and Engineering Science, Shanghai University,
Shanghai 200072, People's Republic of China

Jin-Wu Jiang[*]

Shanghai Key Laboratory of Mechanics in Energy Engineering,
Shanghai Institute of Applied Mathematics and Mechanics,
School of Mechanics and Engineering Science, Shanghai University,
Shanghai 200072, People's Republic of China

Harold S. Park

Department of Mechanical Engineering,
Boston University,
Boston, Massachusetts 02215, USA

**The universality of the models**

The above theoretical models are suitable for any fullerene-based sandwiches. To demonstrate the generality of the model, we perform MD simulations to study $\alpha_{c1}$ and $\alpha_{c2}$ in the MoS$_2$/fullerene/MoS$_2$ sandwich structures as shown in Fig. S1 (a). A Stillinger-Weber potential is used to describe the interatomic interaction within each layer of MoS$_2$ [1] and the AIREBO potential is used for the interatomic interaction in fullerenes, where the LJ potential is used to model the van der Waals interactions among different layers [2]. Fig. S1 (b) shows that $\alpha_{c1}$ approaches to $\frac{1}{\pi}$ in the limit of large area, which further verifies Model I. The critical value $\alpha_{c1}$ increases with the increase of system area, which can be described by Model II with parameters shown in Table. 1.

Figure S1 (c) shows the critical value $\alpha_{c2}$. The MD results can be well described by Model IV with parameters shown in Table. 1. The length of the bending segment, $c$ = 28.1 Å, is larger than the value of 13.11 Å in the GFG sandwich structure, because the bending modulus for MoS$_2$ is much larger than that for graphene. We have thus shown that the analytic models are universal and are applicable to other sandwich structures.

Table 1: The corresponding coefficients

| a | b | c (Å) | $\rho_b$ (eV/Å) | $g_c$ (eV/Å$^2$) |
|---|---|---|---|---|
| 0.0117 | 1.0631 | 28.1 | 0.19 | 0.018 |

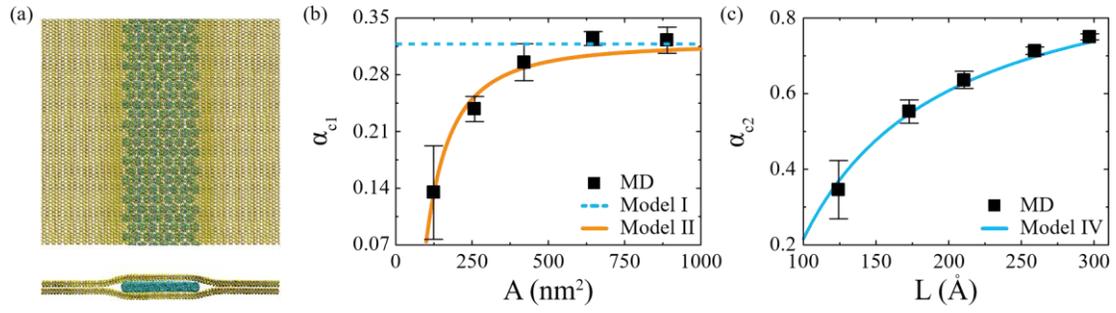

**Fig. S1** Results for MoS$_2$/fullerene/MoS$_2$ sandwich structure. (a) A snapshot of MoS$_2$/fullerenes/MoS$_2$ sandwich structure. (b) The critical value $\alpha_{c1}$ increases with increasing area and saturates at a constant value $\frac{1}{\pi}$. (c) The critical value $\alpha_{c2}$ increases with the increase of the system size.